\documentclass{iaus}
\usepackage{graphicx}
\usepackage{upmath}
\usepackage{amsbsy}

\newcommand{\OIII}{[O{\sc iii}]}
\newcommand{\NII}{[N{\sc ii}]}

\title[The kinematics and binary-induced shaping of PN HaTr~4]  
{The kinematics and binary-induced shaping of PN HaTr~4}

\author[Amy A. Tyndall, David Jones \& Myfanwy Lloyd]  
{Amy A. Tyndall$^1$,
  David Jones$^2$
  \and Myfanwy Lloyd$^1$
 }

\affiliation
{$^1$Jodrell Bank Centre for Astrophysics, University of Manchester, Oxford Road, Manchester, M13 9PL, UK \\email: {\tt amy.tyndall@postgrad.manchester.ac.uk}, {\tt myfanwy.lloyd@manchester.ac.uk}\\[\affilskip]
$^2$European Southern Observatory, Alonso de C\'ordova 3107, Casilla 19001, Santiago, Chile \\ email: {\tt djones@eso.org} \\[\affilskip]
}

\pubyear{2010}
\volume{283}  
\pagerange{x--y}
\jname{Planetary Nebulae: an Eye to the Future}
\editors{A. Manchado, L. Stanghellini \& D. Schoenberner, eds.}
\begin{document}

\maketitle

\begin{abstract}
We present the first detailed spatio-kinematical analysis of the planetary nebula HaTr~4, one of few known to contain a post-Common-Envelope central star system. Based on high spatial and spectral resolution spectroscopy of the \OIII\ nebular emission line, in combination with deep, narrow-band imagery, a spatio-kinematical model was developed in order to accurately determine the three-dimensional morphology and orientation of HaTr~4. The nebula is found to display an extended ovoid morphology with an equatorial enhancement consistent with a toroidal waist - a feature believed to be typical of central star binarity. The nebular inclination is found to be in good agreement with that determined for the binary plane, providing strong evidence that shaping and evolution of HaTr~4 has been influenced by its central binary system - making HaTr~4 one of only 5 planetary nebulae to have had this observationally proven.
\keywords{binaries: close, ISM: kinematics and dynamics, ISM: jets and outflows, planetary nebulae: general}
\end{abstract}

\firstsection 
\section{Introduction}
HaTr~4 is known to contain a non-eclipsing photometric binary central star with a period of 1.71 days (\cite[Bond \etal\ 1990)]{bond90}. On first inspection of the original imagery presented by \cite[Hartl \& Tritton (1985)]{hartl85}, HaTr~4 has the appearance of a classical butterfly bipolar nebula lying in the plane of the sky with twin lobes emanating in an East-West direction. However, deeper H$\alpha$ + \NII\ imagery acquired by D. Pollacco shows faint extensions in a North-South direction indicating that the bipolarity in fact lies perpendicular to that inferred from the original imagery. This deep narrow-band imagery, in combination with high-spectral and -spatial resolution spectroscopy acquired with VLT-UVES, have been used to derive a spatio-kinematic model of HaTr~4 in order to investigate the relationship between the PN and its binary central star (\cite[Tyndall \etal\ 2011a, 2011b)]{tyndall2011a,tyndall2011b}. To date, only four PNe have been observationally confirmed as having been shaped by their central binary stars -- Abell~63 (\cite[Mitchell \etal\ 2007)]{mitchell07}, Abell~41 (\cite[Jones \etal\ 2010)]{jones10}, Abell~65, (\cite[Huckvale \etal\ 2011)]{huckvale11} and NGC~6337 (\cite[Hillwig \etal\ 2010]{hillwig10}, \cite[Garc\'ia-D\'iaz \etal\ 2009)]{garciadiaz09}.

\section{Spatio-Kinematical Modelling}

Using the astrophysical modelling program \textit{SHAPE} (\cite[Steffen \etal\ 2011)]{steffen11}, a spatio-kinematical model was developed in order to reconstruct the nebular morphology of HaTr~4. The nebular expansion velocity was assumed to follow a Hubble-type flow, and the simplest best fitting model was an open-ended ovoid nebular shell waisted by a thick equatorial ring. The model accurately reproduces both the observed velocity profiles and imagery. The model shows that the main nebula does in fact extend in a North-South, and not East-West, direction, and that `butterfly' appearance of the central nebular regions is due to a projection effect.

\section{Results}
HaTr~4 possesses an elongated, axisymmetric morphology with an equatorial enhancement consistent with a nebular ring. The `butterfly' appearance of the central region is shown to result from a line-of-sight inclination effect associated with the enhanced nebular waist, rather than a bipolar structure. The modelling has revealed an age of order 8200 years at a distance of 3 kpc, and an equatorial expansion velocity of 26$\pm$4 kms$^{-1}$  - both fairly typical for PNe. No evidence is found for extended emission or jet-like outflows associated with the PN, such as those found in ETHOS~1 (\cite[Miszalski \etal\ 2011a)]{miszalski11} and The Necklace (\cite[Corradi \etal\ 2011)]{corradi11}. 

The inclination of the nebular symmetry axis is found to be 75$^{\circ}$$\pm$5$^{\circ}$, consistent with the non-eclipsing nature of the binary central star. Further investigation into the central star system performed by Hillwig et al. (in preparation) and Bodman et al. (in press), indicates that the inclination of the central binary plane falls within a similar range to that of the nebula as predicted by binary-induced PN shaping theories (\cite[Nordhaus \& Blackman 2006)]{nordhaus06}. This alignment between the nebular symmetry axis and binary plane provides strong evidence that HaTr~4 has been shaped by its central binary star, making it one of only five PNe to have had this link observationally shown (\cite[Jones \etal\ 2011)]{jones11}.

\end{document}